\documentclass[prb,aps,twocolumn,groupedaddress,floats,showpacs,final]{revtex4-1}
\usepackage{graphicx}
\usepackage{dcolumn}
\usepackage{bm}
\usepackage{color}
\definecolor{blue}{rgb}{0.3,0.3,0.9}
\def\he4{$^4$He}
\def\hee3{$^3$He}
\def\beq{\begin{eqnarray}}
\def\eeq{\end{eqnarray}}

\begin{document}

\title{Giant isochoric compressibility of solid \he4: the bistability of superlcimbing dislocations}

\author{A. B. Kuklov}
\affiliation {Department of Engineering Science and Physics and the Graduate Center, CUNY, Staten Island, NY 10314, USA}

\date{\today}
\begin{abstract}
   A significant accumulation of matter in solid \he4 observed during the superflow events,  dubbed as the giant isochoric compressibility (or the syringe effect),  is discussed within the model of dislocations with superfluid core. It is shown that solid \he4 in a contact with superfluid reservoir can develop a bistability with respect to the syringe fraction, with the threshold for the bias by chemical potential determined by a typical free length of dislocations with superfluid core.  The main implications of this effect are: hysteresis and strongly non-linear dynamical behavior leading to   growth, proliferation and possibly exiting from a crystal of  superclimbing dislocations. Three major channels for such dynamics are identified: i) injection and inflation of the prismatic loops from the boundary; ii) Bardeed-Herring generation of the loops in the bulk; iii)  helical instability of the screw dislocations.    It is argued that the current experiments are likely to be well in this regime.
Several testable predictions for the time and the bias dependencies of the dynamics are suggested.      
\end{abstract}

\maketitle

\section{Introduction}
The superflow through solid \he4 observed first in the UMASS group \cite{Hallock} and then confirmed by other groups \cite{Beamish,Moses} is now at the focus of the experimental and theoretical efforts in the field of superfluidity and quantum crystals. One of the striking features is the syringe effect (or the giant isochoric compressibility) \cite{sclimb}.
In its essence, a solid exhibits the response on external chemical potential applied at a point, practically, the same way as  liquid does --  absorbs or expels a macroscopic fraction of atoms.      

As it has been suggested in Ref.\cite{sclimb}, this effect  can be associated with the so called superclimb of edge dislocations -- the climb supported by the superfluid transport along dislocation core. The unusual feature of this scenario is that the linearized isochoric compressibility of a solid permeated by a network of dislocations with superfluid core is independent of  density of the superclimbing dislocations and is, instead, determined by the dimensionless parameter -- the asymmetry between lengths of superclimbing and non-superclimbing parts. This implies that the effect is strongly non-perturbative, that is, it cannot be treated as a small correction with respect to dislocation density. In particular, as shown in Ref.\cite{sclimb}, the linear isochoric compressibility  of a symmetric network is essentially the same as that of a liquid. Here we show that, even if the initial network is strongly asymmetric in favor of the  non-superclimbing (superfluid) dislocations, there are scenarios still leading to the giant isochoric compressibility. 

\begin{figure}[t]
\centerline{\includegraphics[width=0.9\columnwidth, angle=0]{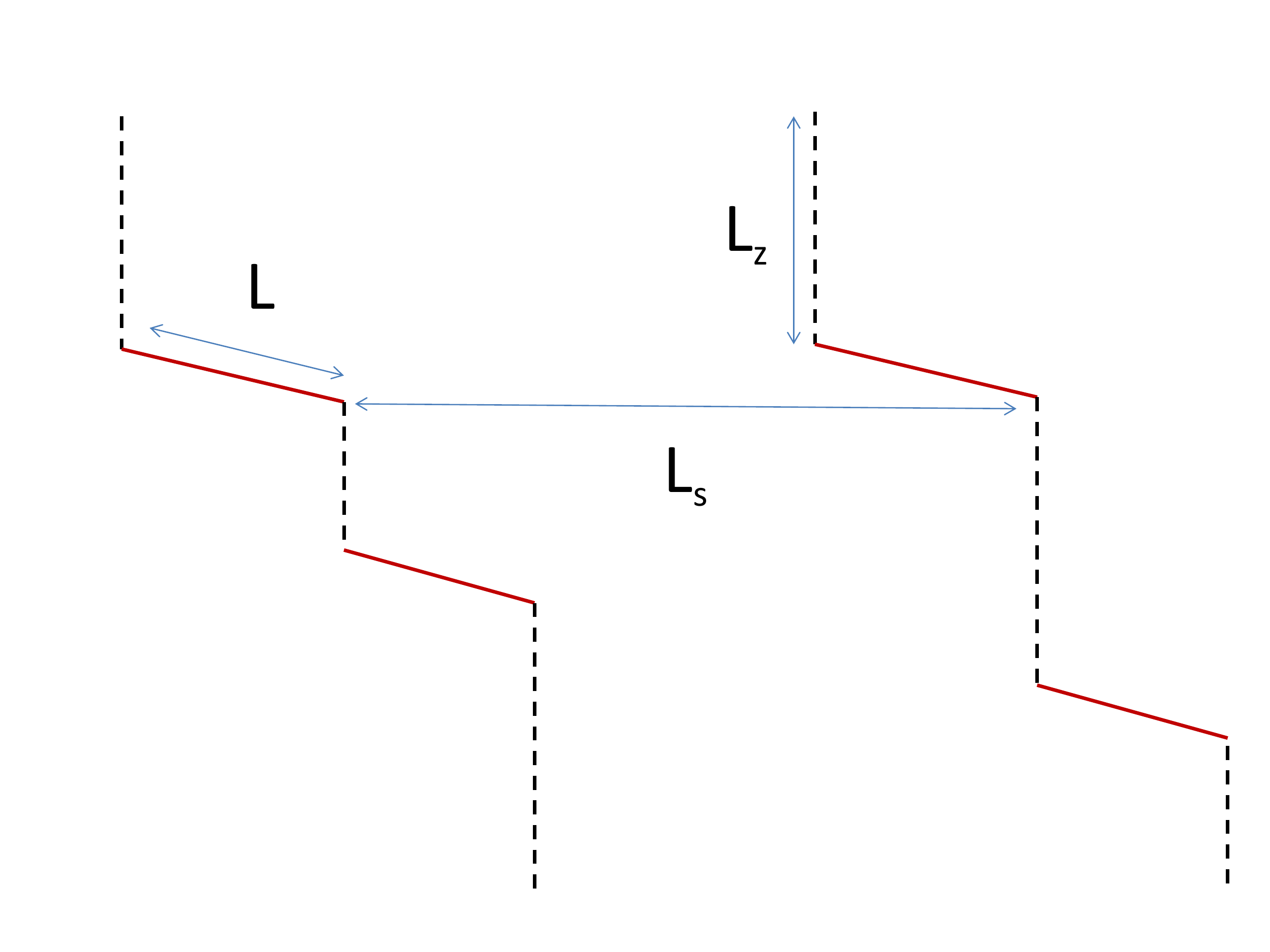}}
\vspace*{-0.5cm}
\caption{(Color online) A forest of screw dislocations containing edge superclimbing segments.  Dashed and solid lines indicate pure screw and pure edge segments, respectively, all characterized by the Burgers vector along the {\it hcp} axis.}
\label{fig_s}
\end{figure}
\begin{figure}[t]
\centerline{\includegraphics[width=0.6\columnwidth, angle=0]{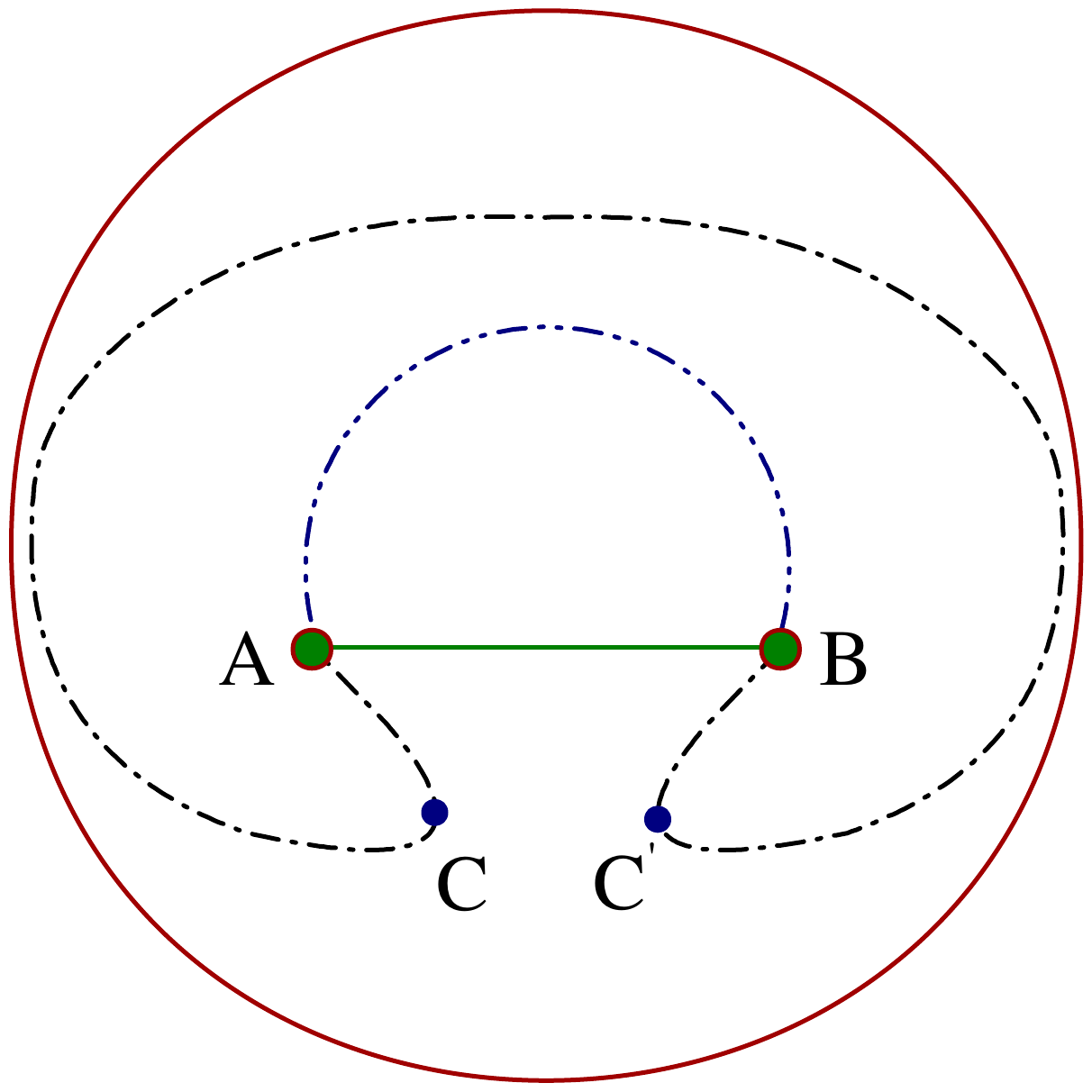}}
\vspace*{-0.5cm}
\caption{(Color online) A prismatic loop (solid line) generated by an edge segment AB, from Fig.~\ref{fig_s}, according to the Bardeen-Herring mechanism: An originally straight edge segment AB (solid horizontal line) bows under the bias (dashed-double-dotted line). Then, further bowing results in the overhangs (dashed-dotted line). The points C,C' in the overhangs approach each other and eventually the whole prismatic loop detaches from the points A,B. 
}
\label{figFR}
\end{figure}
According to the suggestion \cite{sclimb} the vycor "electrodes" are creating a contact between superfluid reservoirs and a preexisting static network of dislocations with superfluid cores. Such dislocations are characterized by Burgers vector along the main symmetry axis and can be of two basic types -- screw \cite{screw} and edge \cite{sclimb} (as well as of the mixed type). The edge part of the network can execute superclimb responsible for the syringe effect. An example of such a network with combined segments is shown in Fig.~\ref{fig_s}. 

There is an alternative to the "preexisting static network" scenario -- a dynamical network which is created and disrupted by the external bias $\mu$.
Here the analysis of the superclimb is extended beyond the linear response considered in Ref.\cite{sclimb}, and it is shown that a segment of a rough superclimbing dislocation is unstable with respect to its unlimited growth, if the bias by chemical potential $\mu$ exceeds a threshold $\mu_c$ which is inversely proportional to  a length $L$   of the segment. In high quality crystals typical value of $L$ can be as large as several $\mu$m or a fraction of mm or even reach a sample size. Thus, the threshold can be {\it macroscopically} small, so that it may well be exceeded by several orders of magnitude in the experiments \cite{Hallock,Beamish,Moses}.   The reason for such a generic situation is that a number of the conducting pathways is $\propto 1/L^2$, so that, in order to detect the superflow, the bias $\mu$ needs to be increased at least as $\mu \propto L^2$ which leads to $\mu >> \mu_c \propto L^{-1}$.

It will also be shown that a straight screw dislocation (which cannot perform superclimb) with superfluid core \cite{screw} can develop a helical instability under the bias so that the edge-like rim is formed and, accordingly, the syringe effect will also be induced.  The threshold for this instability has the same dependence $\propto 1/L$ on  length of the screw dislocation.[Helical screw dislocations have been first observed in silicon at high temperatures \cite{Dash}].

\begin{figure}[t]
\centerline{\includegraphics[width=1.0\columnwidth, angle=0]{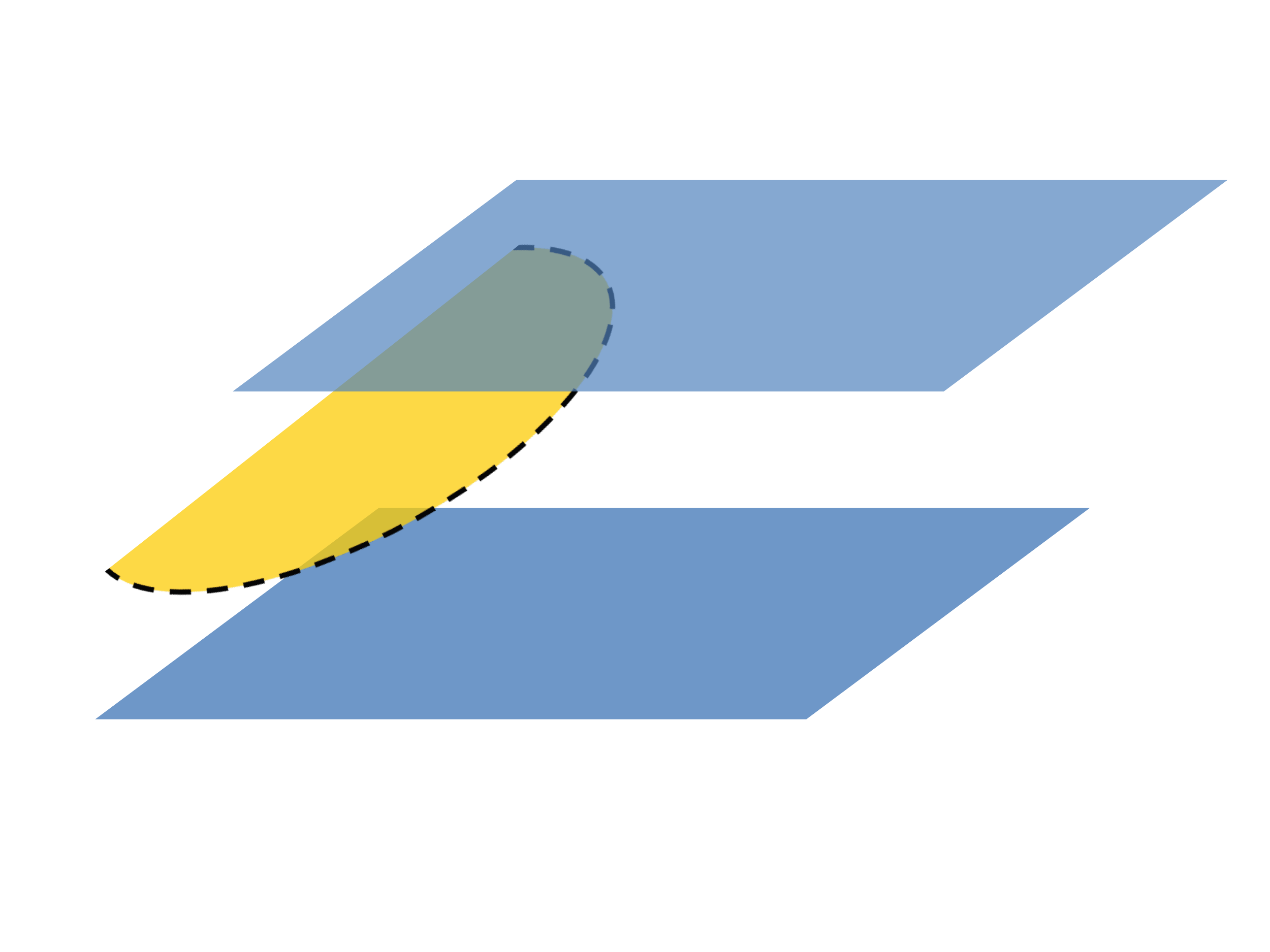}}
\vspace*{-0.5cm}
\caption{(Color online) Sketch of the growing superclimbing dislocation (dashed line). The area to its left indicates  either building of extra plane of atoms or dissolving of the existing plane (not shown) between the upper and lower ones.  }
\label{fig1}
\end{figure}

\begin{figure}[t]
\centerline{\includegraphics[width=0.95\columnwidth, angle=0]{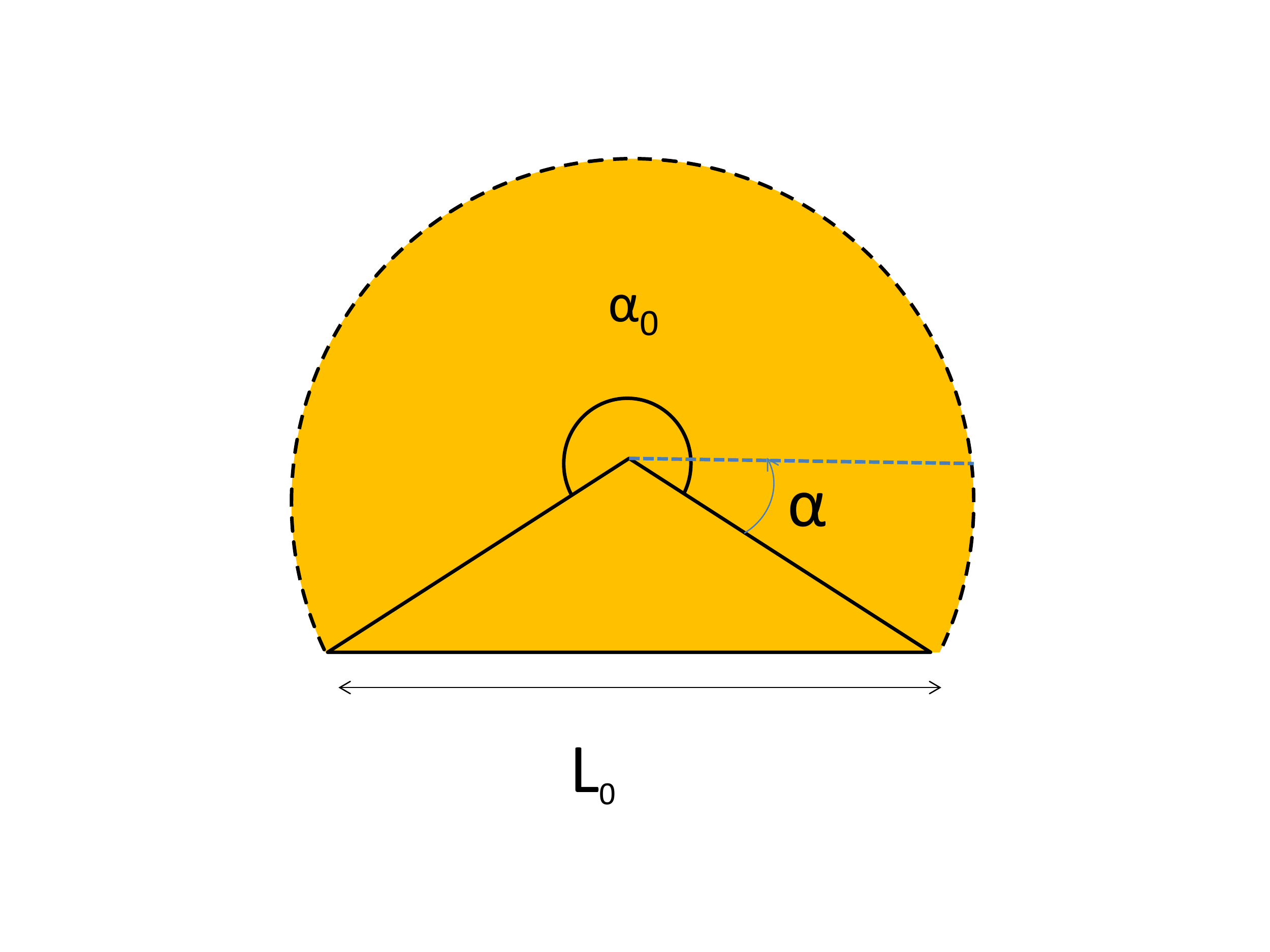}}
\vspace*{-0.5cm}
\caption{(Color online) Geometry of the growing superclimbing dislocation (dashed line) sketched in Fig.~\ref{fig1}. The endpoints of the dislocation are in a contact with the superfluid reservoir.The (solid) area under the curve indicates either the injected extra matter leading to the formation of new  basal layer (actually two of them in {\it hcp}) or a removed part of the existing one.}
\label{fig2}
\end{figure}

The instability has two important consequences: First, rough superclimbing segments of dislocations pinned inside solid \he4 bulk can generate prismatic loops upon bias  $\mu$ in a manner very similar to the Frank-Reed source of gliding dislocation loops under shear stress \cite{FrankReed}. The difference is that the generated loops during the superclimbing instability carry extra matter or vacancies. [In this case the instability should rather be called as Bardeen-Herring \cite{Bardeen}]. A typical diameter of the loop is determined by the original length of the edge segment $L=L_0$.  This process is schematically shown in Fig.~\ref{figFR}.  Second,
 superclimbing dislocations existing at a solid-vycor boundary  can proliferate into the bulk upon applying the bias, so that a percolating network of superfluid pathways is created even if it didn't exist originally. This process is shown in Figs.~\ref{fig1},\ref{fig2}.  Both effects are symmetric with respect to the sign of the bias $\mu$. In one case an additional matter is injected into the solid in the form of parts of extra basal planes, and in the second -- existing basal planes are being dissolved, that is, vacancies are being injected instead. 

The presented analysis is conducted at the level of a single dislocation. It ignores how pinning by \hee3 impurities or crosslinks with other dislocations may affect the dynamics of the instability. It is clear that the instability may also result in the dislocations exiting the solid from its edges. Several growing segments may also merge or recombine. These processes  as well as the interaction between superclimbing and basal plane gliding dislocations are also not considered. In some sense the analysis presented here is limited by low density of superclimbing dislocations so that there is some reasonably long time during which the dynamics can be treated within an approximation of a single dislocation segment.
This situation is different from the linearized approach \cite{sclimb} where the main assumption was that a typical distance between superclimbing segments is of the same order as a typical length of the each segment. The single loop strongly non-linear dynamics considered here relies on a different limit -- that is, a typical distance between superclimbing segments is much larger than $L_0$.

The injection of the dislocations from the vycor-solid boundary and the Bardeen-Herring type loop generation as well as the helical instability of screw dislocations  result in the syringe effect. The dynamics of the instabilities, however, turn out to be different: while in the case of the boundary instability the injected dislocation can grow to sizes far exceeding its original length $L=L_0$, in the case of the Bardeen-Herring type instability the generated loop radius $R$ is of the order of $L_0$ -- it is the number of loops that is changing.  The helical screw dislocation can also generate loops in a manner similar to that proposed in Ref.\cite{Dash}. [The detailed study of this effect in the context of the superfluid core will be conducted elsewhere].         




\section{Growth instability of rough superclimbing dislocation} 
Let's consider one segment of a superclimbing dislocation of some  initial  length $L=L_0$. Such a segment can be at a crystal boundary or be a part of the superfluid network. If the bias $\mu$ is applied, the segment will bow due to an extra matter delivered through its ends.  Such bowing occurs in the basal plane while the Burgers vector $b$ is perpendicular to the plane, that is, along the high symmetry axis. This process is schematically shown in Fig.~\ref{fig1} as an arc protruding between two basal layers.

At this point, let's specify what the bias $\mu$ is. An increase of chemical potential of the superfluid reservoir $\mu_l$ either by applying pressure \cite{Hallock,Beamish,Moses}  or through the Fountain effect \cite{Hallock} creates a difference $\mu = \mu_s - \mu_l <0$ between chemical potentials of the solid $\mu_s$ and the liquid. As a result, an additional matter can be injected into the solid in a form of growing pairs of basal plane layers. The boundary of one pair of layers is the superclimbing dislocation.   If $\mu >0$, an existing pair of layers is being dissolved. Its boundary is also a superclimbing  dislocation. The added or removed part of the planes is shown by a colored solid area under the arc in Fig.\ref{fig1}.      
 
It is important to realize that imposing any finite value of $\mu$ (that is deviation from the equilibrium value) results, strictly speaking, in the instability.
This can be understood from simple energy balance:  the energy gain due to the bowing $\delta E_b = |\mu \delta N|$, with $\delta N$ being a number of atoms delivered through the core to support the bowing, always exceeds the energy $\delta E_{\rm cr} \propto (L-L_0)$ due to the core length increase from $L_0$ to $L$ for large enough $L$ because $\delta N$ is given by the area swept during the bowing.  Thus, for large enough $L$ the energy gain due to the bowing always dominates the energy loss due to the length increase.
However, in the limit $\mu \to 0$ the (meta)stability is protected   by a macroscopic energy barrier. This barrier vanishes if $|\mu| $ exceeds a threshold $ \sim 1/L_0$ so that the absolute instability develops. As the estimates provided later show, the actual experiments \cite{Hallock,Beamish,Moses} appear to be well in this regime.

Let's estimate the threshold value $\mu_c$ for the bias. The energy of the dislocation per its unit length is given by shear modulus $G$ and Burgers vector $b$ as $\epsilon_c \approx Gb^2  /4\pi$, so that $\delta E_{\rm cr} \approx \epsilon_c (L-L_0)$. The energy gain $\delta E_b$ scales by the area $\sim |\mu| (L/b)^2$ swept by the bowing dislocation. Thus, equating one to the other gives $\mu_c$ as
\beq
\mu_c \sim  \frac{Gb^4}{ L_0}.
\label{inst}
\eeq      
The value of $b$ in Eq.(\ref{inst}) is the Burgers vector along the {\it hcp} axis, that is, $b=\sqrt{8/3} a$ where $a$ is the interatomic distance $a \sim 3.5$\AA. As was found in the simulations \cite{screw}, this dislocation splits into two partials with $b \to b/2$ and the fault in between. Thus, effectively, $b$ is reduced by a factor of 2 so that the actual threshold (\ref{inst}) becomes lower by a factor of about $2^4$. In the following discussions I will ignore this peculiarity of the structure, which can only modify the numerical coefficient without affecting the dependence $ \mu_c \propto 1/L_0$. Accordingly, in all the following estimates the value of $b$ will be taken as $b \approx a \approx 3.5$\AA ~ and the core splitting will be ignored.

The above relation can be supported by a more quantitative analysis.
In the quasi-static situation, so that $\mu$ is the same over the whole dislocation length,  the bowing dislocation takes the shape of circular arc characterized by the base $L_0$ and the angle $\alpha$ as indicated in Fig.~\ref{fig2}. 
At small $|\mu|$ the center of the circle is outside the crystal so that the circle radius $R>> L_0$ and $\alpha_0 \to 0$. As the arc grows, $\alpha_0$ eventually reaches $\pi$  and $R$ decreases to $R=L_0/2$ and then $\alpha_0 \to 2\pi $ so that the center of the circle enters the crystal and $R$ starts growing to become $R>>L_0$ .  
The total energy  of such a configuration can be represented as 
\beq
E=\frac{Gb^2}{4\pi}R\cdot \alpha_0 - \frac{|\mu|}{b^2} R^2\cdot (\alpha_0 - \sin \alpha_0) 
\label{arc}
\eeq
where the radius $R$ of the arc is determined by $L_0$ and the angle $\alpha_0$ as $L_0=2 \sin(\alpha_0/2) R$. This expression indicates that the dislocation is absolutely unstable toward inflation $R \to \infty, \, \alpha_0 \to 2\pi$ for arbitrary small $|\mu|$ simply because the second term is negative and can dominate at large enough $R$. There is, however, an energy  barrier to overcome before the instability develops unless $|\mu|$ exceeds some critical value.  Let's consider a specific situation when the end points of the dislocation are pinned by a contact with a superfluid reservoir so that $L_0$ is fixed. Then, the energy (\ref{arc}) becomes a variable of the angle $\alpha_0$ only:
\beq
E(L_0, \alpha_0)= E_0 \frac{\alpha_0 \sin(\alpha_0/2) - \tilde{g}\cdot (\alpha_0 - \sin \alpha_0)}{\sin^2(\alpha_0/2)}, 
\label{Ed}
\eeq
where $E_0 \approx Gb^2L_0/8\pi$, $\tilde{g}\equiv \pi |\mu|  L_0/( b^4 G)$.
 As the analysis of the function (\ref{Ed}) indicates, for $\tilde{g}<<g_c=0.5$ there is a metastable minimum at $\alpha \approx 4\tilde{g} <<1$ followed by a maximum at larger $\alpha_0$. At $\tilde{g}=g_c$ both,the minimum and the maximum, coincide at $\alpha_0=\pi$ which is the inflection point indicating ending of the metastability domain, so that at $\tilde{g}>g_c$ the dislocation becomes unstable toward unlimited inflation $\alpha_0 \to 2\pi$. As can be seen, the  threshold $\tilde{g}=0.5$, that is,
\beq
\mu_c= \frac{Gb^4}{2\pi L_0}
\label{inst2}
\eeq
 is consistent with the estimate (\ref{inst}).

In a general situation one should expect a distribution of the lengths $L_0$ so that some segments remain in a metastable equilibrium and some are overcritical. In what follows such a distribution will be ignored and it will be considered that there are $M$ segments of some length $L_0$ in a solid affected by the bias $\mu$. 
 Typical values utilized in the flow experiments \cite{Hallock}, when the syringe effect was observed, are in the range
$\mu =5\cdot 10^{-4} - 5\cdot 10^{-3} $K (which corresponds to $0.001-0.01$J/g in units of Ref.\cite{Hallock}). This implies that  the  lengths of the critical segment $L_0\geq 1-5\mu$m (for a typical $G\approx 100$bar and $b\sim 3.5$\AA).  The expected density of dislocations in high quality \he4 crystals is at the level of $\sim 10^4-10^6$cm$^{-2}$ as found in Ref.\cite{Balibar}, which implies that the actual lengths $L_0$ of free segments are about a factor of 10-100 longer than the above estimate. In other words, the experimental range of $\mu$ used in Ref. \cite{Hallock} appears to be well above the threshold (\ref{inst}) (or (\ref{inst2})). At this point it should be mentioned that the dislocation density values \cite{Balibar} are more relevant to the glide effect than to the superclimb. Nevertheless, this order of magnitude estimate can be used as a figure of merit.

\subsection{Helical instability of the screw dislocation with superfluid core}
As found in {\it ab initio} simulations \cite{screw}, screw dislocation in solid \he4 with Burgers along the {\it hcp} axis has a superfluid core. If this dislocation is straight, there are no edge-type segments on it, and, therefore, it cannot perform the superclimb. Here it will be shown that, if such a dislocation is biased by chemical potential similarly to the edge segment discussed above, it will become unstable toward forming a helix with its axis parallel to the original orientation of the dislocation. Such a helix has the edge-type rim and, thus, it can be a cause of the syringe effect. 

Let's consider a screw dislocation of length $L$ oriented along the z-axis (that is, the {\it hcp} axis). Then, a  position of the core can be described in the cylindrical coordinates by the radial distance $r(z)$ from its original position $r=0$ (in units of $b$) as well as by the azimuthal  angle $\theta (z)$. The energy consists of two terms: the work done $ \mu \Delta N$  by the bias $\mu$ to accumulate some amount of matter $\Delta N$ due to creating the edge-type rim and the energy $\sim \epsilon_c$  due to the core length increase :
\beq
E_s= \int_0^{L}dz  \left\{\frac{ \mu \gamma_s r^2}{2}\partial_z \theta  + \frac{\epsilon_c}{2} [ (\partial_z r)^2 + r^2 (\partial_z \theta)^2 ]\right\}, 
\label{Lscr}
\eeq
where it was taken into account that the additional matter per unit length of the core is $d \Delta N/dz  = \gamma_s r^2 \partial_z \theta /2$, with $\gamma_s=\pm 1$ being the chirality (handedness) of the dislocation. In other words, the total amount $\Delta N$ is given by the projection of the helix on the basal plane times the number of the complete turns. The "sign" of the matter accumulation depends on $\gamma_s$: if the screw is right handed, $\gamma_s=1$, and the helix is right handed, $ \partial_z \theta >0$, the solid mass increases, that is, $\Delta N >0$. Similarly, $\Delta N>0$ for both the screw and the helix  being left handed ($\gamma_s =-1,\, \partial_z \theta <0$). Conversely, the amount of the syringe matter becomes negative if the chiralities are opposite to each other. Eventually, it will be seen that the sign of the syringe fraction does not depend on the screw chirality and is   solely determined by the sign of the bias $\mu$ as $\Delta N \sim - \mu$.

 As a specific choice of the boundary condition, let's presume that this dislocation is pinned at its both ends, that is,  $r(0)=r(L)=0$.
Then, the variation with respect to $\theta$ gives the equation 
\beq
 \partial_z \theta =- \frac{\mu \gamma_s}{2\epsilon_c },
\label{hel}
\eeq
where the boundary condition is taken into account.
Its substitution back to Eq.(\ref{Lscr}) results in the effective energy of the dislocation as
\beq
E= \int_0^{L}dz  \left\{- \frac{(\mu  r)^2}{8\epsilon_c}+ \frac{\epsilon_c}{2}  (\partial_z r)^2 \right\}. 
\label{Lscr3}
\eeq
 This expression features an instability toward unlimited growth of $r$. At small $\mu$, similarly to the case of the edge dislocation, the solution $r=0$ is a metastable one. As Eq.(\ref{Lscr3}) indicates, there is a difference with the edge dislocation case -- the screw  does not show any linear response of bowing in the limit $\mu \to 0$. There is, however, a threshold $\mu_s$ such that at $|\mu | > \mu_s$ the absolute instability toward $r \to \infty $ develops. In order to find how $\mu_s$ depends on $L$ and $\epsilon_c$ it is enough to perform elementary estimates: $r <<L$ changes on the scale of $L$ so that the total elastic energy is $ \sim \epsilon_c r^2 /L $. As long as the bias energy $ \sim \mu^2 L r^2 /\epsilon_c $ becomes of the same order, the solution $r=0$  becomes absolutely unstable. Thus, 
\beq
\mu_s \approx \frac{\epsilon_c}{L}\approx \frac{Gb^4}{L},
\label{mu_s}
\eeq    
which is essentially the same condition as (\ref{inst}).
At the threshold the helix is described by the total angle $|\alpha | \approx \mu_s L/\epsilon_c \approx 1$, and at $|\mu| >>\mu_s$ this angle becomes $|\mu|/\mu_s >>1$.

\subsection{Collective elastic effect in a  bulk network of superfluid dislocations}
The injection of extra matter (or vacancies) under the bias $\mu$ is limited by the compression elastic modulus $K$ of a sample, so that the system stabilizes at some finite density of extra matter delivered through superclimb. The above estimate (\ref{inst}) (or (\ref{inst2})) obtained for a single dislocation does not take  into account this effect and implies that an inflating loop can reach a sample size. In reality, the generation must stop after the overall density change compensates for the bias $\mu$.

 Let's presume that the extra matter resides in 
$M$  dislocation segments of length $L_0$ which bowed by an amount $y $ each. Such bowing results in the extra matter $\Delta N \sim L y  M/b^2$ added to (or subtracted from) the solid. There is the corresponding compression energy increase    $E_e \approx K (\Delta N/N)^2 \Omega/2$, where $N$ stands for the total number of particles in the bulk of the volume $\Omega$ affected by the injection. The chemical potential energy gain and the energy loss due to the core length increase are $\sim \mu L y M/b^2 $ and $\sim Gb^2 y^2/L$, respectively. Thus, the total energy change as a function of $y,M,L$ becomes
\beq
E\approx -  \frac{LM \mu}{b^2}y + \left[\frac{Gb^2M}{L}+ \frac{K  L^2b^2M^2}{\Omega}\right] \frac{y^2}{2},
\label{Etot}
\eeq 
where the dimensionless numerical coefficients are omitted.
If $|y| <<L$, the value of $L$ can be set to $L\approx L_0$ and the minimization in $y$ gives 
\beq
y\approx \frac{L_0^2/(Gb^4)}{1+ KL_0^3M/(\Omega G)}\cdot \mu.
\label{y}
\eeq
As mentioned above, this solution is actually a metastable one, which, however, is protected by exponentially long waiting time in the limit $ \mu \to 0$.
The bowing determined by (\ref{y}) corresponds to the syringe fraction $\Delta N/N \sim MLy b/\Omega$ which for the case depicted in Fig.~\ref{fig_s}, where $M \sim \Omega/(L_s^2 L_z)$ can be written as
\beq
\frac{\Delta N}{N}\approx  \frac{L_0^3/(L_s^2L_z)}{1+ KL_0^3/(GL_s^2L_z)}\cdot\frac{\mu}{Gb^3}.
\label{frac}
\eeq
If $L_s \sim L_z \sim L$, that is, for a uniform network of the superclimbing dislocations, this fraction becomes
\beq
\frac{\Delta N}{N} \approx \frac{\mu }{(K+G)b^3},  
\label{comp}
\eeq
which constitutes the giant isochoric compressibility \cite{sclimb}. In the limit $L^3_0 << L^2_s L_z$, the syringe fraction becomes reduced within the linearized approach as
\beq
\frac{ \Delta N}{N} \approx \frac{L_0^3}{L_s^2L_z} \frac{\mu }{Gb^3} << \frac{\mu }{Gb^3},  
\label{comp3}
\eeq

If the bias $\mu$ exceeds the threshold (\ref{inst}), the bowing of the edge segments cannot be treated in the linear approximation anymore.~
 In order to find the syringe fraction in a generic situation one can use Eq.(\ref{Etot}) where the substitute $y \sim L$ is made and, instead of $L$, the fraction $ N_1 \sim L^2/b^2$ generated by one segment is used as a variable. Then, Eq.(\ref{Etot}) becomes:
\beq
E\approx - |\mu| M N_1 + Gb^3 M \sqrt{N_1}+ \frac{K  b^6 M^2 }{2\Omega} N_1^2 ,  
\label{Etot2}
\eeq 
where $N_1$ is taken as a positive value featuring either extra matter delivered to ($\mu <0$) or taken out from ($\mu > 0$) the solid.

 For small enough $M$ this function of $N_1$ features a maximum at $N_1=N_{mx} \sim (Gb^3/|\mu|)^2$ and then a stable minimum at $N_1=N_{eq} $ where
\beq
N_{eq} \approx \frac{|\mu | \Omega}{MKb^6}.
\label{Lm}
\eeq
This minimum corresponds to the syringe fraction $\Delta N/N = N_1 M b^3/\Omega$, that is,
\beq
\frac{\Delta N}{N} \approx \frac{\mu }{Kb^3},  
\label{comp2}
\eeq
where the condition $N_{mx} << N_{eq}$, that is,
\beq
|\mu| \geq \mu_b \approx \left(G^2 K \frac{M}{\Omega}\right)^{1/3} b^4
\label{Nmxmn}
\eeq 
must hold.  In the case of $M$ bulk segments distributed uniformly in, e.g., the situation depicted in Fig.~\ref{fig_s}, the condition (\ref{Nmxmn}) becomes $|\mu| >  \left(G^2 K/L_s^2L_z\right)^{1/3} b^4$. Thus, if $L_0$ is the smallest length scale and $|\mu|$ obeys (\ref{inst}), the system is guaranteed to be unstable, with the equilibrium (\ref{Lm}) to be determined by the bulk elastic energy $E_e$. The fraction (\ref{Lm}) corresponds to the limit $|\mu| >> \mu_b$.

In the case of the bulk structure shown in Fig.~\ref{fig_s} this fraction $N_1$ is residing in several prismatic loops $N_{lp}$ of a radius $R \sim L_0$ generated by each edge segment.  This number can be estimated as $N_{eq}b^2/L_0^2$, that is,
\beq
N_{lp}  \approx \frac{|\mu| L_s^2L_z}{Kb^4 L_0^2} > \left(\frac{GL_s^2L_z}{KL^3_0}\right)^{2/3} >> 1, 
\label{Req}
\eeq
 where the condition (\ref{Nmxmn}) as well as that $L=L_0$ is the smallest distance among $L_s,L_z,L$ in Fig.~\ref{fig_s}  are taken into account.

Thus, in the overcritical regime the equilibrium fraction (\ref{comp2})  is always of the same order as in the liquid -- even if the linearized response predicts much smaller values (\ref{comp3}).

\subsection{Collective elastic effects due to the boundary instability} 
In the case of the vycor-solid boundary, the analysis should be performed separately because seeds of the unstable dislocations are residing at the boundary. Accordingly,
$M$ in this case is rather a surface than the bulk quantity. Then, in the estimate of the bulk deformation energy the affected volume becomes $\Omega \sim L S$, where $S$ stands for the area of the vycor-solid boundary. Accordingly, the extra fraction of the injected matter/vacancies is $\Delta N\sim L^2 M/b^2$ and $N\sim \Omega/b^3$ so that
\beq
\frac{\Delta N}{N} \sim \frac{L M b}{S}, 
\label{surfra}
\eeq
 and the elastic energy takes the form
\beq
E_e \approx \frac{ KM^2 b^2L^3}{2S}.
\label{vs}
\eeq 
This dependence $\propto L^3$ should be contrasted with the elastic term $\propto N_1^2 \propto L^4$ in the case of the bulk instability as represented in Eq.(\ref{Etot2}). The total energy in terms of $N_1$ (that is, the extra matter due to one segment) becomes
 \beq
E\approx - |\mu| M N_1 + Gb^3 M \sqrt{N_1}+ \frac{K  b^5 M^2 N_1^{3/2}}{2S} .
\label{Etot3}
\eeq 
This form as a function of $N_1$ (or the segment length $L$) can also have two extrema -- a maximum followed by the minimum as $N_1$ grows. It happens when
\beq
|\mu | \geq \mu_s \approx \left(\frac{GKM }{S}\right)^{1/2}  b^4.
\label{Nmxmn2}
\eeq 
This condition for the surface bistability should be compared with the bulk one (\ref{Nmxmn}). The value of $M/S$ is determined by typical distances between the boundary segments along the basal plane $r_b$ and along the hcp axis $r_z$ as $ M/S \approx 1/r_b r_z$. Thus, if $L_0 < \sqrt{r_b r_z}$, the condition (\ref{Nmxmn2}) is guaranteed to be satisfied as long as the instability condition (\ref{inst}) holds. If $|\mu| >>\mu_s$, the equilibrium is determined by the first and the last terms in the energy (\ref{Etot3}). It corresponds to the typical equilibrium length (obtained from the minimization of $E$ in (\ref{Etot3}) with respect to $N_1 \sim L^2$) as
\beq
L=L_{eq} \approx \frac{|\mu| r_b r_z }{ K b^4}.
\label{Leq}
\eeq
It is interesting to note that this length becomes of the order of a sample size ($\sim $ 1 cm) for the smallest values of the bias $\sim 5\cdot 10^{-4}$K  used in Ref.\cite{Hallock} ( or $\mu \sim 10^{-3}$J/g in units of Ref.\cite{Hallock}) , if $r_b,r_z$ are of the order of the vycor diameter $\sim 1$mm. This value, however, drops quite fast with the product $r_br_z << 1$mm$^2$.   In other words, if there are only few seeds of superclimbing dislocations at the solid-vycor boundary, the instability guarantees that the new pathways will reach the other electrode. Conversely, if there are many such seeds, the elastic energy increase due to the injection will stop the syringe effect close to the boundary. 

Generically, a system of superclimbing dislocations can feature two minima with respect to the syringe fraction. These minima are separated by a barrier, and, therefore, the hysteresis phenomenon should be anticipated with respect to the bias as long as the condition (\ref{Nmxmn}) (or (\ref{Nmxmn2})) is satisfied. 

\subsection{Renormalization of the chemical potential}
The instability of a single superclimbing dislocation is eventually stabilized by the increase of the bulk elastic energy due to finite density of the injected matter (or vacancies). This corresponds to the renormalization of the difference of chemical potentials $\mu$ between solid and liquid from its initial value to zero.
This renormalized value $\tilde{\mu}$ can be obtained from the expressions of the total energy, Eqs.(\ref{Etot2},\ref{Etot3}), as $\tilde{\mu} = \partial (E/M) /\partial N_1$. Keeping in mind the symmetry $\mu \to - \mu$ let's consider $\mu < 0$, that is, that the potential of the liquid in vycor is higher than in the solid so that extra atoms enter the solid. In the case of the bulk segments (as in Fig.~\ref{fig_s}) the differentiation of the energy (\ref{Etot2}) results in
\beq
\tilde{\mu} = \tilde{\mu}(N_1)=\mu + \frac{Gb^3}{2}N_1^{-1/2} + \frac{Kb^6M}{\Omega} N_1 ,
\label{mu_bu}
\eeq   
   and in the case of the vycor-solid boundary Eq.(\ref{Etot3})  gives
\beq
\tilde{\mu} =\tilde{\mu}(N_1)= \mu + \frac{Gb^3}{2}N_1^{-1/2} + \frac{3Kb^5M}{4 S} N^{1/2}_1 .
\label{mu_b}
\eeq   

 There are two roots of $\tilde{\mu}=0$. At small enough $M$ ( as presented in Eqs.(\ref{Nmxmn},\ref{Nmxmn2}) and guaranteed by the generic condition (\ref{inst})) the first one $N_1 \approx (Gb^3/(2 \mu))^2$ corresponds to unstable equilibrium, and the second one describes  stable one. It should be mentioned that the equilibrium characterized by small bowings (that is, $N_1 \to 0$) is not captured by Eqs.(\ref{mu_bu},\ref{mu_b}) written for already large bowings $y \sim L_0$. Thus, for all practical purposes this minimum can be viewed as corresponding to the energy $E=0$ reached at $N_1=0$. 

As discussed above, the practical values of $\mu$ corresponds to the situation when $N_1$ is to evolve from the unstable toward the stable equilibrium. This what is called above as the overcritical regime $|\mu| > \mu_c$, Eq.(\ref{inst}). As will be seen, the overcritical dynamics  exhibit strongly non-linear features before $N_1$ approaches the vicinity of the stable equilibrium.
 
\subsection{Liquid-gas type transition and hysteresis}
It is useful to look on Eqs.(\ref{Etot2},\ref{Etot3}) from a different perspective. Small bowing of dislocations corresponds to $N_1 \to 0$, that is to zero energy $E=0$. There can exist another equilibrium solution $\partial E/\partial N_1=0$ characterized by  finite $N_1$. Thus, there is a value of chemical potential $|\mu|=\mu_I$ at which two phases $N_1\approx 0$ and $N_1$ finite have the same energies. This can be interpreted as a point of first order phase transition.  For the case of the bulk system,Eq.(\ref{Etot2}),  
\beq
\mu_I = 1.5  \left(G^2 K \frac{M}{\Omega}\right)^{1/3} b^4.
\label{Ist}
\eeq
At smaller values of $|\mu|$ the second solution for $N_1$ becomes metastable and at $|\mu| =\mu_{sp}$, where
\beq
\mu_{sp} = 2^{-1/3}\mu_I \approx 0.794 \mu_I,
\label{sp}
\eeq
 it vanishes. Thus, $|\mu |=\mu_{sp}$ corresponds to the spinodal, that is, to the point where the bistability (and hysteresis) vanishes. 

Similar situation occurs for the case of the interface, Eq.(\ref{Etot3}).   The transition occurs at $|\mu|=\mu_{Is}$, where
\beq
\mu_{Is} =  \sqrt{2} \left(G K \frac{M}{S}\right)^{1/2} b^4,
\label{Ist2}
\eeq
and the hysteresis vanishes at $|\mu|=\mu_{sps}$, where
\beq
\mu_{sps} = \frac{\sqrt{3}}{2}\mu_{Is} \approx 0.866 \mu_{Is}.
\label{sp2}
\eeq

The transition is not characterized by any underlying symmetry, and, to some extent, resembles Ist order liquid-gas transition, where density exhibits a jump. However, there is also a significant difference. 
The energies (\ref{Etot2},\ref{Etot3}) contain essentially a non-analytical term $\sim \sqrt{N_1}$ determined by the geometrical nature of dislocations.
This term is always dominant at small $N_1$ and is the reason for the energy barrier. Thus, in contrast to the standard liquid-gas transition,
the syringe effect does not have a critical point where the first order transition ends.

\section{Dynamics of the instability at the vycor-solid boundary}    
Now let's consider the dynamical aspects of the evolution of the syringe fraction. In this work, the focus is on the dynamics of the edge segments. The dynamics of the helical instability will be analyzed elsewhere.

A single loop dynamics during the instability stage is characterized by  a short ballistic period during which phase slips and dissipation can be ignored. Then, as the superfluid velocity along cores reaches some terminal value, phase slips induce dissipation which is strongly non-linear in the velocity. Finally, once length of a growing segment approaches equilibrium value (determined by the largest root of $\tilde{\mu}=0$ in Eq.(\ref{mu_b})), the linearized dynamics sets in. The analysis is conducted within the assumption that distance between dislocations is large enough so that there is enough time for a segment to grow to a length $L$ which is  much larger than the initial length $L_0$. Within this approach other growing loops are taken into account self-consistently through the renormalization of the chemical potential given by the last term in Eqs.(\ref{mu_bu},\ref{mu_b}). 
 Practically, this means distances $\sim 10^{-3}-10^{-2}$cm for dislocation densities $10^6-10^4$cm$^{-2}$.

\subsection{Ballistic growth from the boundary}
  
The analysis will be conducted for an almost circular loop of radius $R$ (that is, $\alpha_0 \approx 2\pi$ in Fig.~\ref{fig2}). In general, the parametrization of the loop should include deviations of its shape from circle. This can be achieved by introducing $R$ as a function of polar angle $\alpha$ (as defined in Fig.~\ref{fig2}) and time $t$. 
The Lagrangian  can be written as
\beq
{\cal L}= \int_0^{2\pi} d\alpha  \frac{- \dot{\phi} \sigma R^2}{2} - E_1
\label{L}
\eeq
where the first term is the Berry contribution in units $\hbar=1, b=1$ , and $\sigma = \pm 1$ is to be chosen depending on whether the matter is injected to ($\sigma =+1$) or from ($\sigma =-1$ ) the solid; the term $ E_1 $ is the energy of one loop which takes into account the elastic energy, that is, the last term in Eq.(\ref{Etot3}), attributed to one loop. In addition, the kinetic energy of the flow along the core $ \frac{\rho_s(\partial_{\alpha} \phi)^2 }{2  dl /d\alpha}$, where $\phi$ is the superfluid phase and $\rho_s$ stands for the superfluid stiffness and $ \frac{dl}{d\alpha}=\sqrt{R^2 + (\partial_{\alpha} R)^2}$, should be taken into account. Thus,
\beq
E_1 = \int_0^{2\pi} d\alpha \left[\frac{\rho_s(\partial_{\alpha} \phi)^2 }{2  dl /d\alpha}   - |\mu| N_1 + \epsilon_c \frac{dl}{d\alpha} \right] + \kappa N^{3/2}_1,
\label{Le}
\eeq
where $\kappa \approx Kb^5 M/S, \, \epsilon_c= Gb^3/4\pi $ and $N_1$ as the amount of extra matter absorbed by one loop is $N_1 = \int_0^{2\pi} d\alpha R^2/2$.
As discussed above, the system is invariant with respect $\mu \to - \mu$. Thus, without loss of generality the value of $\mu$ can be taken negative and $\sigma =+1$ so that $N_1$ describes extra matter added to the solid. 
There is also the boundary condition indicating that the dislocation is in a contact with the superfluid reservoir at its two endpoints, that is, $\phi(\alpha=0)=\phi(\alpha=2\pi)=\phi_R$, where $\phi_R$ is the phase of the reservoir, which can be set to zero.

 Within the simplified approach deviations from the circular shape can be ignored, that is, $\partial_{\alpha} R=0$. Then, a certain minimal assumption must be made about the spatial dependence of the phase $\phi$ along the dislocation core.  Given the symmetry of the problem, Fig.~\ref{fig2}, and because the total current through the loop must be zero in the syringe regime, the current along the core $ \sim \partial \phi /\partial \alpha$ must be antisymmetric with respect to $\alpha=\pi $ .Thus, the lowest non-trivial angular harmonic satisfying this requirement as well as the boundary condition is 
\beq
\phi = \phi_0(t) \sin\left(\frac{\alpha}{2}\right).
\label{ans}
\eeq 
 Then, a substitution of this ansatz into Eqs.(\ref{L},\ref{Le}), after performing explicit integration and variation in $\phi_0$, gives
\beq
\frac{ dR^2}{dt}= \frac{\pi \rho_s}{8}  \frac{\phi_0}{R}.
\label{var_phi}
\eeq 
 This equation is essentially the statement of the continuity: the flux of matter through two ends of the growing loop controls the  rate of the loop  area change.
The variation of the action in $R$ gives 
\beq
 \frac{d\phi_0}{dt}=  \bar{\mu},
\label{R}
\eeq 
where
\beq
 \bar{\mu}= \frac{\pi |\mu|}{2} + \frac{\pi \rho_s \phi_0^2}{32 R^3}  - \frac{\pi\epsilon_c}{2R} - 3 \pi^{3/2} \kappa R.
\label{barmu}
\eeq
For $|\mu|$ exceeding the threshold (\ref{inst}) and when $R$ is yet far from the equilibrium, the dominant time dependence is determined by the first two terms in $\bar{\mu}$. Then, the solution of Eqs.(\ref{var_phi},\ref{R},\ref{barmu}) can be looked for in the form $\phi_0= A t, R= B t^\nu$ with some unknown parameters $A,B,\nu$. This gives 
\beq
\phi_0(t)= \frac{3\pi \mu}{4} t, \,\, R(t)=\left(\frac{9\pi^2\rho_s|\mu|}{128}\right)^{1/3} t^{2/3}.
\label{phi}
\eeq
These dependencies describe the balistic stage of the syringe effect far from the equilibrium.
The accumulated fraction is given by Eq.(\ref{surfra}) where the role of $L$ is played by $R$ from Eq.(\ref{phi}), that is,
\beq
\frac{\Delta N}{N} \propto (\rho_s |\mu|)^{1/3} t^{2/3},
\label{DNbal}
\eeq
 which corresponds to the local pressure variation $ \sim K |\Delta N|/N $. 
These fractional powers are specific to the dislocation superclimb and, thus, their experimental observation would be a "smoking gun" for the effect. The question, though, is how long this ballistic stage can last.

 During the ballistic stage the flow velocity grows  in time until some terminal velocity $V_T$  is reached (at the dislocation end points). Then, frequent  phase slips take place which convert kinetic energy of superflow into excitations. Thus, as an order of magnitude estimate, a typical time for the phase slip $t_{ps}$ can be taken from the ballistic stage --- how  long it takes to accelerate the flow from zero to, say, $V_T \sim 100 $m/s. The velocity profile along the growing loop is determined by the phase (\ref{ans}) as $V(t,\alpha) = \partial_\alpha \phi /(mR)$, that is,
\beq
V(t,\alpha)= \frac{\phi_0(t)}{2mR(t)} \cos(\alpha/2),
\label{vel}
\eeq     
where $m$ is \he4 atomic mass.
Using the solution for $\phi_0$ and $R(t)$ in this equation, one finds 
 \beq
V(t, \alpha)=  V_0\cos(\alpha/2), \, \, V_0=\frac{\hbar}{mb} \left(\frac{t}{\tau_b}\right)^{1/3}.
\label{vel2}
\eeq     
in the standard units, where the time scale $\tau_b$ is given by
\beq
\tau_b^{-1}=\frac{3\pi \mu^2b}{4\rho_s}\approx 2.4 \frac{ \mu^2 m b}{\hbar^3  n^{(1d)}_s} ,
\label{tau}
\eeq
 and $ n^{(1d)}_s\approx m\rho_s $ is the superfluid linear density along the dislocation core. Using its value  $ n^{(1d)}_s \sim 1$\AA$^{-1}$ found in simulations of screw dislocation \cite{screw} and $b\sim 3.5$\AA (so that $\hbar/mb \approx 50$m/s), the terminal speed $V_T$ (at $\alpha=0,2\pi$) is reached at time $ t_T \approx  \tau_b$,
which for the  lowest $\mu$ value used in Ref.\cite{Hallock} gives $\sim 1$ ms. At this moment it looks unlikely that the available time resolution allows observing the time dependence during this stage. There is, however, an initial portion of the syringe effect which is  on the experimental time scale (of minutes) \cite{Hallock,Beamish} is accumulated in a jump-like manner right after the bias $\mu$ is imposed (or removed). This quantity will be discussed later.

Close to the stable equilibrium point $R\to R_{eq}$  (determined by $\dot{\phi}=0$), where
\beq
R_{eq}=\frac{|\mu|}{6\sqrt{\pi} \kappa},
\label{Req_b}
\eeq
which is essentially Eq.(\ref{Leq}),
Eqs.( \ref{var_phi},\ref{R}) can be linearized $R=R_{eq} + \xi $ with $|\xi| << R_{eq}$. The corresponding dynamics is oscillatory:
\beq
\ddot{\xi} + \omega_\mu^2\xi=0,\,\, \omega^2_\mu= \frac{27\pi^{7/2} \rho_s \kappa^3}{4\mu^2}.
\label{osc}
\eeq
Thus, $\omega_\mu $ scales as $\propto 1/|\mu|$. These oscillations, however, can take place only if their amplitude is small so that the velocity of the flow along the core remains much smaller than the terminal velocity. Thus, detecting them presents a significant challenge. It should also be mentioned that the phase slips in the ohmic regime at finite temperature $T$ (see below) may make the oscillations overdamped.

\subsection{Dissiptaive stage}

At the experimental time scale of minutes \cite{Hallock,Beamish} the dynamics is dominated by diffusive (dissipative) processes. The nature of these processes is not exactly known. One possible scenario is that quantum phase slips assisted by thermal processes in the superflow along dislocation cores are responsible for the dissipation. The dependence of the flow rate vs bias \cite{Hallock} is consistent with the picture of the phase slips in Luttinger liquid containing a weak link  \cite{Kane,Borya_Kolya}. 
At the same time, the origin of a significant temperature dependence \cite{Hallock,Beamish} observed in the non-linear regime is not fully understood.
In this situation, a phenomenological approach should be used in order to describe the loop inflation. Specifically, in the ballistic dynamical equation (\ref{R}) the l.h.s. can be rewritten
in terms of the velocity amplitude $V_0$ in Eq.(\ref{vel}) as $\phi_0 \to mR V_0$.  Then, the effective friction rate $  \gamma(T,V_0) V_0$, with some friction coefficient $\gamma$ depending on $V_0,T$, should be added to the flow acceleration $dV_0/dt$ in Eq.(\ref{R}). This transforms Eq.(\ref{R}) into     
\beq
& &\frac{dV_0}{dt}+ \gamma V_0 \approx \frac{\bar{\mu}}{mR}
\label{RV} \\
\bar{\mu}&\approx& \frac{\pi |\mu|}{2} + \frac{\pi m^2 \rho_s V_0^2}{32 R}  - \frac{\pi\epsilon_c}{2R} - 3 \pi^{3/2} \kappa R,
\label{Vdis} 
\eeq
where the Bernoulli pressure is now expressed in terms of the velocity rather than the phase. 

The growth rate of the loop area $\sim \dot{R^2}$ is determined by $V_0$. The actual relation (stemming from the continuity equation) is exactly the same as in Eq.(\ref{var_phi}) where, however, the phase is now expressed in terms of the velocity $V_0$:
\beq
\frac{\pi dR^2}{dt}= 2 \rho_s V_0,
\label{Rate2}
\eeq 
where it is understood that $V=V(\alpha =2\pi)= - V(\alpha=0)$ so that the matter is delivered symmetrically from the both ends of the growing loop of radius $R >>L_0$.

Let's now make a choice for $\gamma$. According to the quantum phase slip scenarios \cite{Kane,Borya_Kolya}  energy of 1D superflow is converted into excitations of the Luttinger liquid. Generically, the quantum effects assisted by thermal excitations are characterized by power law dependencies of the phase slip rate $\tau_{ps}^{-1} \propto T^\zeta$ with some $\zeta > 0$ determined by the Luttinger liquid parameter $g$. At zero $T$ the flow velocity $V$ controls the dissipation. Within the weak link situation a phase jump by $\sim \pi$ occurs at microscopic distances across the link which is reasonable to take as $\sim b$.  This jump is accompanied by energy transfer $\sim V$ between the link and the Luttinger liquid. Thus, $V_0$ plays the role of temperature so that the dependence on $V_0$ should be characterized by the same exponent $\tau_{ps}^{-1} \propto |V_0|^\zeta$, with   the crossover taking place at some $V_0=V_T \approx Tb$. Thus, as a single equation the friction rate in Eq.(\ref{RV}) can be represented
as
\beq
\tau^{-1}_{ps} \sim \gamma V_0 = \gamma_0 {\rm Im}[bT+ i V_0]^{\zeta},
\label{gamma}
\eeq
   with some coefficient $\gamma_0$. In the weak-link scenario $\gamma_0$ is determined by frequent phase slips occurring at the location of the link. Thus, the associated time constant $ \propto \gamma_0^{-1}$ can be much shorter than a typical time-scale set by the period of Debye frequency in solid \he4. This may essentially eliminate the ballistic stage for practical durations of the experiments. So, below the estimate for the loop inflation will be obtained under this assumption, that is, that the ballistic stage is too short to produce any significant syringe effect.

The power $\zeta$ can be empirically related to the power $p$ observed in the flow rate vs bias dependence $ |V| \propto |\mu|^p$ in Ref. \cite{Hallock}
as $ \gamma V_0 \sim |\mu|^p$ so that 
\beq
\zeta = p^{-1}.
\label{zeta}
\eeq
 According to Ref.\cite{Kane} $\zeta = 2/g -1 $ and the self-consistent result \cite{Borya_Kolya_prv} gives $\zeta = 2g -1$ . In what follows the power $p$ will be used as a quantity measured directly in the experiment \cite{Hallock}. This power was found to vary in the range $0.25<p < 0.5$ .

For $|V|<<V_T=Tb$ Eq.(\ref{gamma}) implies ohmic regime
\beq
 \gamma V_0 = \gamma_0 {\rm Im}[bT+ i V_0]^{1/p} \to  p^{-1} \gamma_0 (bT)^{p^{-1} -1} V_0.
\label{gammaT}
\eeq

Eqs.(\ref{gamma},\ref{gammaT}) will be used below in the analysis of the loop dynamics in the long-time limit where the inertial part in Eq.(\ref{RV}) can be omitted. 
Then, far from the equilibrium at low $T$ the dynamics is dominated by the first term in the brackets of Eq.(\ref{Vdis}).
Thus, 
\beq
V_0\approx \left(\frac{|\mu|}{m \gamma_0 R}\right)^{p} ,
\label{VR}
\eeq
in the non-linear regime (\ref{gamma}), and
\beq
V_0\approx \frac{p(bT)^{1- p^{-1}}}{m \gamma_0 R} |\mu|,
\label{VR_ohm}
\eeq
 in the ohmic regime (\ref{gammaT}).

These expressions must be used in Eq.(\ref{Rate2}). Accordingly, in the non-linear regime the loop radius obeys
\beq
\frac{dR^2}{dt} \approx 2 \rho_s \left(\frac{|\mu|}{m \gamma_0 R}\right)^{p},
\label{non-lin}
\eeq
which implies $R \propto  |\mu/\gamma_0|^{p/(2+p)} (t\rho_s)^{1/(2+p)}$, or for the syringe fraction
\beq
\frac{\Delta N}{N} \propto R \propto    \left(\frac{|\mu|}{\gamma_0}\right)^{\frac{p}{2+p}} ( \rho_s t)^{\frac{1}{2+p}}.
\label{DN_b}
\eeq
As Eq.(\ref{VR}) indicates, the flow speed actually drops with time as $V \propto t^{- 1/(2+p)}$. Thus, eventually, the non-linear regime must change to the ohmic one characterized by
\beq
\frac{dR^2}{dt} \approx 2 \rho_s \frac{p T^{1-p^{-1}}}{m \gamma_0 } \frac{|\mu|}{R}, 
\label{ohm}
\eeq
  which gives 
\beq
\frac{\Delta N}{N} \propto R \propto  T^{\frac{1- p^{-1}}{3}} \left(\frac{\rho_s |\mu| t}{\gamma_0}\right)^{\frac{1}{3}}.
\label{ohmDN}
\eeq

Eqs.(\ref{DN_b},\ref{ohmDN}) are obtained under the assumption that the system is far from the equilibrium. If, however, it approaches the equilibrium, the last term in the brackets 
of Eq.(\ref{Vdis}) becomes important (with the second and the third ones still being irrelevant). This term stabilizes the system at the equilibrium radius $R_{eq}$, Eq.(\ref{Req_b}). Close to the equilibrium the dynamics becomes linear in the deviation $|R_{eq} - R| << R_{eq}$. As mentioned above, the time dependence would become either dissipative at high $T$ or oscillatory as in Eq.(\ref{osc}). 

At this point it should be mentioned that the stabilization of the instability may also happen due to the dynamical rather than due to static equilibrium. Specifically, if $R_{eq}$ exceeds a system size, the stabilization is to be achieved by the balance of growing new loops and the loops exiting the sample. This picture essentially depends on sample geometry and size and will not be discussed here.

\subsection{The   jump in the syringe fraction due to the ballistic inflation}
As discussed above, the ballistic stage may lead to a jump of the accumulated syringe fraction right after the bias is applied (or removed).
Let's estimate this fraction, first, for $T=0$. In the dynamical  equation  (\ref{RV}) the dissipative part can be ignored as long as $|\dot{V}| >> \gamma_0 V^{1/p}$.
Using the ballistic solution (\ref{vel2}) in this estimate, the limiting time becomes
\beq
t_{bal} \propto \frac{\tau_b^{\frac{1-p}{1+2p}}}{\gamma_0^{\frac{3p}{1+2p}}} \propto \frac{\rho_s^{\frac{1-p}{1+2p}}|\mu|^{-\frac{2(1-p)}{1+2p}}}{\gamma_0^{\frac{3p}{1+2p}}}, 
\label{tbal0}
\eeq 
where the definition (\ref{tau}) of $\tau_b$ is used. A substitution of it into the ballistic fraction, Eq.(\ref{DN_b}), gives the jump
as
\beq
\frac{|\Delta N|}{N} \propto \frac{\rho_s^{\frac{1}{1+2p}}}{\gamma_0^{\frac{2p}{1+2p}} |\mu |^{ \frac{1-2p}{1+2p}}}, \,\,\,\, |\mu| >\mu_c.
\label{max_0}
\eeq     
The value of $p$ was found in Ref.\cite{Hallock} to be below 0.5. Thus, the ballistic jump is a {\it decreasing} function of the bias, provided it exceeds the threshold for the instability and the jump itself does not exceed the equilibrium syringe fraction $\propto |\mu|$.  [In this case, the last term in $\bar{\mu}$, Eq.(\ref{Vdis}), should be taken into account
which will change the ballistic solution (\ref{vel2})]. However, as mentioned earlier, the friction "amplitude" $\gamma_0$ in the denominator may actually suppress the jump below the experimental resolution. 

Let's now consider finite $T$. Comparing the acceleration rate with the thermal phase slips in Eq.(\ref{RV}) the ballistic evolution takes place (before it is interrupted by the ohmic regime) as long as $t$ is shorter than the smallest of either the ohmic dissipation time $\gamma^{-1}_0 T^{1-1/p}$ or the time when the terminal velocity $V=bT$ is reached. Clearly, at very small $T$ and $p<1$ the latest dominates, which from Eq.(\ref{vel2}) follows as $t\approx t_{T} \propto \tau_b T^3$. Then, at longer times the evolution becomes essentially the same as that at $T=0$ and leads to the estimates (\ref{tbal0},\ref{max_0}).   However, at the experimental values of $T$ and large $\gamma^{-1}_0$, the estimate $t< t_{T} \approx \tau_b (\gamma^{-1}_0 T^{1-1/p} )^3 $ is more appropriate for the time limiting the ballistic evolution.  Then, a substitution of $t_{T}$ into Eq.(\ref{DNbal}) gives the jump as
\beq
\frac{|\Delta N|}{N} \propto \frac{ \rho_s T^{2(1-p^{-1})}}{\gamma_0 |\mu|}.
\label{max_bal}
\eeq    
This dependence should be considered in the limit $\gamma_0 \to \infty$, that is, that the maximum typical time for the phase slips $\sim  \gamma^{-1}_0$ is below
 $(T/T_0)^{2+1/p}\tau_b$, where $T_0\sim 1K$ is a typical temperature  corresponding to the velocities $\sim 100$m/s. At this point, it is worth mentioning that the jumps of the syringe fraction have been observed in Ref.\cite{Beamish}. To what extent these can be interpreted in terms of the ballistic stage remains to be seen.

\section{The Bardeen-Herring type instability}
While a dislocation injected from crystal edge to the bulk can grow up to $R_{eq}$ which is much larger than its initial size ( or even as large as  sample size),  a finite superclimbing segment inside a solid, e.g., in the case shown in Fig.~\ref{fig_s}, can generate loops only of a size of the order of its initial length. According to the Bardeen-Herring mechanism \cite{Bardeen} originally considered for gliding dislocations and known as Frank-Reed instability \cite{FrankReed}, an initially straight segment bows under the bias, and eventually the overhangs are created, Fig.~\ref{figFR}. These overhangs merge together so that a circular (prismatic) loop of a radius $R_L$, which is of the order of initial length $L_0$ of the straight segment, is created. This process is cyclic and is characterized by  time $t_{FR}$ needed for the loop to grow until the overhangs (C,C' in Fig.~\ref{figFR}) merge together so that the loop becomes separated from the main network.       
At this point what happens to this loop is not important -- it can,e.g., diffuse away or merge with newly created loops.

An estimate for this time can be obtained from Eq.(\ref{non-lin}) in the non-linear regime or from Eq.(\ref{ohm}) in the ohmic regime, where the time $t_{FR} $ is found as a function of the loop radius $R $ reaching the length of the order of the  original segment length $L_0$. It is natural to assume that this time $t_{FR}$ is much shorter than the experimental time $t$, so that many loops are generated by one segment before the equilibrium $\tilde{\mu}=0$ is reached. Thus, the accumulated fraction (far from the equilibrium) can be written as $\Delta N/N \propto t/t_{FR}>>1$. Thus, the syringe rate $d \Delta N/dt$  becomes 
\beq
\frac{ d \Delta N}{dt} \propto \frac{1}{t_{FR}} \propto \frac{\rho_s |\mu |^{p}}{\gamma^p_0L_0^{2+p}}
\label{rate_nn}
\eeq
in the non-linear regime and
\beq
\frac{ d \Delta N}{dt} \propto \frac{1}{t_{FR}} \propto \frac{\rho_s T^{1- p^{-1}} |\mu |}{\gamma_0L_0^3}
\label{rate_ohm}
\eeq
in the ohmic one. After the bulk accumulated fraction approaches the equilibrium value (\ref{comp2}) the constant rates (\ref{rate_nn},\ref{rate_ohm}) transform into the exponential diffusive slowing down.
In contrast to the boundary instability in this case, the Bardeen-Herring mechanism is inherently dissipative and no oscillations are to be anticipated.   

For short loops the Bardeen-Herring cycle can occur in the ballistic regime. In this case the time $t_{FR}$ can be estimated from Eq.(\ref{phi}) as
 $t_{FR} \propto L_0^{3/2} (\rho_s |\mu|)^{-1/2}$. Thus the rate becomes
\beq
\frac{ d \Delta N}{dt} \propto \frac{1}{t_{FR}} \propto \frac{(\rho_s |\mu |)^{1/2}}{ L_0^{3/2}}.
\label{rate_ohm}
\eeq

\section {Discussion}
Solid \he4 with finite density of superclimbing dislocations in a contact with superfluid reservoir is found to be, in general, characterized by bistability with respect to the syringe fraction. This feature is due to the interplay between three contributions: i) the chemical potential energy gain due to a transfer of atoms between two phases -- solid and superfluid; ii) the energy of  the deformation of dislocations needed to accommodate the transfer; and iii) the collective elastic energy. The control parameter of the system are chemical potential and the density of the dislocations. At low densities of dislocations the two fractions are very different from each other and, therefore, the transition between them can be viewed as strongly Ist order transition with significant hysteresis. There is a similarity between this and  liquid-gas transitions, with the exception of no critical point in the first case.

It is highly likely that the syringe effect observed in Refs.\cite{Hallock,Beamish} is  essentially in the overcritical regime. In this regime the equilibrium syringe fraction is given by the liquid type isochoric compressibility despite that the linearized response may predict much smaller values. 

The are two major scenarios for the instability.
First, the vycor-solid boundary can be a source of the superclimbing dislocation loops entering the bulk. Its dynamics is characterized by the ballistic and dissipative stages which can be ohmic or strongly non-linear in the flow velocity. Each regime is characterized by specific powers of the bias and time at the initial stages of the evolution, Eqs.(\ref{DNbal},\ref{DN_b},\ref{ohm}). 

Second, there is also an option for the bulk syringe effect where the accumulated fraction is distributed evenly through out the bulk.
In its turn, the bulk scenario can proceed in two ways -- through  the Bardeen-Herring generation of the prismatic loops or through the helical  instability of screw dislocations. The accumulated fraction in the case of the Bardeen-Herring instability is determined by constant rate dependencies (\ref{rate_nn}), (\ref{rate_ohm})  in the non-linear and ohmic regimes, respectively. The non-linear regime  (\ref{rate_nn}) turns out to be showing the same type of the dependence of the syringe rate on the bias as the flow rate through the sample observed in the Ref.\cite{Hallock}.    In this regard, it is worth mentioning a possibility that the flow through solid may not actually be taking place through a static  network of dislocations percolating between both vycor electrodes. Instead, the loops generated during the Bardeen-Herring cycles may eventually be moving between two vycor electrodes. These loops are mobile due to the superflow along their rim and can serve as "vehicles" transporting the mater across a sample.  Center of mass speed $V_{cm}$ of such a loop is locked to the speed of the superflow $V$   along its rim by a simple geometrical relation $V_{cm} \sim V b/R$ stemming from the matter conservation.   
 Experimental studies of the actual bias-time-temperature dependencies of the syringe fraction and rates are needed to see if any of the above scenarios take place. 

One of the key questions to answer is about the nature of the $T$-dependence observed in the non-linear regime of the flow rate \cite{Hallock}. Similar dependence was also observed in a different setup \cite{Beamish}. Eqs.(\ref{DN_b},\ref{ohmDN},\ref{rate_nn},\ref{rate_ohm}) contain the superfluid density $\rho_s$  in the corresponding powers as overall factors.   
To what extent the observed temperature dependence can be attributed to these factors remains to be seen. 
 One possibility could be that the superfluidity along the cores is strongly affected by structural excitations of dislocation -- kinks \cite{Aleinikava_2012} and jogs -- so that as $T$ increases these excitations suppress the overall superfluid density $\rho_s$ in the cores and, thus, reduce the total flow rate. It should also be mentioned that superclimbing dislocation does not fit exactly into the paradigm of Luttinger liquid because its excitation spectrum is not linear in the momentum \cite{sclimb}. To what extent this feature may modify the results (\ref{gamma},\ref{gammaT}) is an open question too.

The "smoking gun" evidence for truly superfluid flow would be the detection of the ballistic jump in the syringe fraction which is a {\it decreasing } function of the bias. Some jumps have been observed in Ref.\cite{Beamish}. Thus, their detailed study is of crucial importance. 

The main assumption of this work is that density of superclimbing dislocations is  low and a sample size is large enough so that the equilibrium for the generated loops is achieved at typical sizes smaller than sample size. If this condition is not satisfied, as it could be the case for very small samples \cite{Moses}, a completely different scenario may take place: the conducting network may be created by dislocations proliferating directly between the electrodes.  In this case the actual dynamics may be controlled by changing number of the conducting pathways, that is, balanced by the pathways creation and exiting from a sample.       

\section{Acknowledgements} I thank Boris Svistunov and Nikolay Prokof'ev for fruitful discussions. This work was supported by  the NSF grant PHY1314469.

\end{document}